# Plausibler Vorschlag für eine deterministische Wellenfunktion

Petra Schulz

Theodor-Francke-Weg 65, D-38116 Braunschweig

**Kurzfassung**
Es wird ein deterministisches Drehvektor-Modell für Photonen vorgestellt, das auch für Teilchen geeignet ist. Bei einer Kreisbewegung um eine Achse hat die deterministische Wellenfunktion *a* die folgende Form

$$a = \omega_s \, r \, e^{\pm i \omega t}.$$

Dabei bedeuten $\omega_s$ entweder die axiale oder die skalare Spin-Kreisfrequenz (letztere ist proportional der Masse), *r* Radius der Kreisbahn (auch Amplitude einer sich später durch Wechselwirkung ergebenden Schwingung aus zwei entgegengesetzt kreisenden verschmolzenen Photonen), $\omega$ Kreisbahn-Frequenz (ein Maß für die Bahngeschwindigkeit) und *t* Zeit. Das „+" vor dem imaginären *i* bedeutet eine rechtshändige und das „-" eine linkshändige Rotation. Eine Wechselwirkung tritt ein, wenn sich Teilchen einschließlich der Photonen durch Stoßprozesse begegnen und dabei verschmelzen. Dann werden die Wellenfunktionen einfach nur addiert. Sie müssen nicht normiert, geschweige denn renormiert werden.

**Abstract**
A deterministic axial vector model for photons is presented which is suitable also for particles. During a rotation around an axis the deterministic wave function *a* has the following form

$$a = \omega_s \, r \, e^{\pm i \omega t}.$$

$\omega_s$ is either the axial or scalar spin rotation frequency (the latter is proportional to the mass), *r* radius of the orbit (also amplitude of a vibration arising later from the interaction by fusing of two oppositely circling photons), $\omega$ orbital angular frequency (proportional to the velocity) and *t* time. „+" before the imaginary *i* means a right-handed and „-" a left-handed rotation. An interaction happens if particles (including the photons) meet themselves through collision and melt together. The wave functions are then only added simply. They do not have to be normalized or renormalized.

## 1. Einleitung

Ich möchte den Extrakt meiner letzten fünf Arbeiten über das Thema Photon, die ich auf DPG-Tagungen vorgetragen habe [1] bis [5], konzentriert zusammenfassen und etwas erweitern. Das soll in Form einer einzigen Gleichung geschehen. Diese Gleichung paßt locker auf eine Postkarte, die dazu notwendigen Erklärungen leider nicht.

Ich werde dabei an die Erfahrungen anknüpfen, die wir aus Lissajous-Figuren ziehen können. Ich werde neue Spielfiguren zulassen, indem ich mich nicht nur auf senkrecht stehende Schwingungen beschränken werde.

Und im übrigen: die heutigen Ergebnisse mit fast deterministischen Atom-Photon-Laserexperimenten [6] schreien förmlich nach einer deterministischen Wellenfunktion.

Einige neue Grundannahmen der Physik werden sich nicht vermeiden lassen.

## 2. Zur Kinematik
### 2.1 Alle Bewegungen auf Kreisbahnen

Alle natürlichen Bewegungen der Teilchen und Photonen vollziehen sich auf Kreisbahnen mit gleichförmiger Geschwindigkeit. Also wird damit das 1. Newtonsche Gesetz reformiert. (Lineare Bewegungen gibt es zwar auch, aber diese ereignen sich mit ungleichförmiger Geschwindigkeit wie bei einer Schwingung.)

Die Hypothese der natürlichen Kreisbahnen wird z. B. durch die theoretische Arbeit von Boardman und Mitarbeitern indirekt untermauert [7]. Zum Thema elliptische Bewegung komme ich später. Ganz wichtig zum Thema Kreisbewegung ist das Bohrsche Atommodell, das hiermit wiederbelebt werden soll.

### 2.2 Modell der komplexen Zahlen wird erweitert

Für gleichförmige Bewegungen auf Kreisbahnen bietet sich natürlich das Modell der komplexen Zahlen an. Da sich das bisherige Modell nur in einer Ebene abspielt, muß die Erweiterung der einzigen Drehachse auf drei senkrecht stehende Achsen eingebaut werden, s. Abb. 1. Somit entsteht ein räumliches Axialvektormodell (Drehvektormodell). Die Beschreibung der Erweiterungen steht im Anhang.

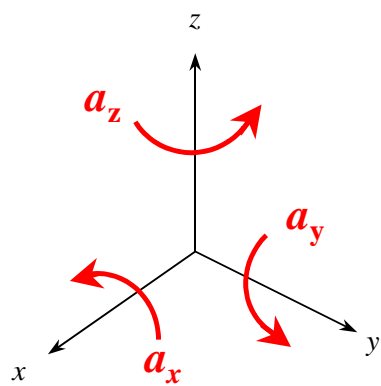

Abb. 1: Die drei axialen Komponenten in einem rechtshändigen kartesischen Koordinatensystem

### 2.3 Bahngeschwindigkeit der Photonen
Die Bahngeschwindigkeit der Photonen im Vakuum vollzieht sich mit Vakuumlichtgeschwindigkeit *c* [3], auch wenn die Photonen einfache bis vielfache Grundfrequenzen einer „Schwingung" tragen. „Rote" und „blaue" Photonen fliegen z. B. gleich schnell.

Mit diesen Annahmen aus Kapitel 2 ist die Kinematik (die reine Kreisbewegung) erledigt. Was nun fehlt, ist die Beschreibung der Körper einschließlich Masse und Spin der Teilchen und Photonen.

### 3. Zum Körper der Photonen und Teilchen
### 3.1 Figur der Photonen und Teilchen
Als Figur der einfachsten Photonen wie etwa das Photon des Elektrons werden rotierende Kügelchen postuliert sowie auch für die einfachsten Teilchen.

### 3.1 Spinrichtung
Die Spinrichtung der Photonen ist so wie bei den Teilchen. Es werden drei senkrecht stehende Drehachsen für die Spins angenommen mit entsprechendem links- und rechtshändigen Spin-Drehsinn [1]. In Abb. 2 sind nur rechtshändige Spins dargestellt.

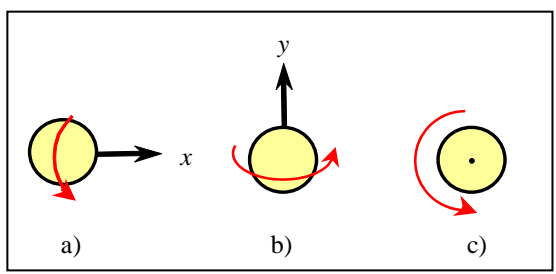

Abb. 2: Die drei Spinrichtungen stehen senkrecht aufeinander und zeigen in die Richtung der Achsen eines kartesischen Koordinatensystems. In c) ragt die Spitze der z-Achse aus der Papierebene heraus.

### 3.2 Spinbetrag ist variabel
Der Spinbetrag der Teilchen/Photonen ist variabel, wie wir es von den Frequenzen her gewohnt sind. Der Spin wird im folgenden durch die Spin-Kreisfrequenz ausgedrückt und nicht durch Quantenzahlen, die unnütz in die Irre führen könnten. Meine Kritik zu den Quantenzahlen und den Hinweis auf experimentelle Untermauerung habe ich in [1] und [4] einfließen lassen.

### 3.3 Axiale und skalare Spin-Kreisfrequenz
Wenn die **axiale Spin-Kreisfrequenz** (oder drehvektorielle Spin-Kreisfrequenz) verwendet werden soll, wird sie fett gedruckt: $\boldsymbol{\omega}_s$. Sie hat drei rechtshändige und drei linkshändige Komponenten. Zu den einzelnen Komponenten werden im folgenden die Richtungen der Drehachsen mit angegeben. Die Pfeilrichtung aus der Papierebene heraus wird als „•", die Gegenrichtung als „x" gekennzeichnet. Die drei rechtshändigen Komponenten, dargestellt in Abb. 2, lauten

$$\boldsymbol{\omega}_s(\rightarrow),\ \boldsymbol{\omega}_s(\uparrow),\ \boldsymbol{\omega}_s(\bullet)$$

und die drei linkshändigen

$$\boldsymbol{\omega}_s(\leftarrow),\ \boldsymbol{\omega}_s(\downarrow),\ \boldsymbol{\omega}_s(\text{x}).$$

Ein kurzes Beispiel zur axialen Spin-Kreisfrequenz wird in Kapitel 6.1 vorgestellt.

Die **skalare Spin-Kreisfrequenz** $\omega_s$ wird in normaler Strichstärke gedruckt und ist die Summe der Beträge der Spin-Kreisfrequenz-Komponenten, also

$$\omega_s = |\boldsymbol{\omega}_s(\rightarrow)| + |\boldsymbol{\omega}_s(\uparrow)| + |\boldsymbol{\omega}_s(\bullet)|$$
$$+ |\boldsymbol{\omega}_s(\leftarrow)| + |\boldsymbol{\omega}_s(\downarrow)| + |\boldsymbol{\omega}_s(\text{x})|. \quad \text{(Gl. 1)}$$

Bei den meisten Experimenten sind die Spinrichtungen der Photonen nicht bekannt, weil keine spinpolarisierten Photonen verwendet werden. Dann wird automatisch nur die skalare Spin-Kreisfrequenz gemessen.

### 3.4 Dimension der Frequenzen
Wichtig ist es nun zu wissen, daß sowohl die axiale wie skalare Spin-Kreisfrequenz als auch die sonst übliche Frequenz $\nu$ eine neue Dimension haben müssen: **Spinumdrehungen/s**.

### 3.5 Umrechnung von Masse aus der Frequenz
Die skalare Spin-Kreisfrequenz $\omega_s$ kann auch durch die Masse ausgedrückt werden und zwar nach der Gleichung für den maximalen Energieübertrag (Spin-Kreisfrequenz-Übertrag) des Photons auf das Elektron beim Compton-Effekt [2].

$$\hbar\,\omega_s = \tfrac{1}{2}\,m\,c^2 \qquad \text{(Gl. 2)}$$

$\hbar$ ist das Plancksche Wirkungsquant geteilt durch $2\pi$, $m$ ist der Masse-Zuwachs, $c$ ist die Vakuum-Lichtgeschwindigkeit. Gleichung 2 ist bis auf den Faktor ½ mit der De-Broglie-Beziehung identisch.

### 4. Vereinfachte deterministische Wellenfunktion
Die vereinfachte deterministische Wellenfunktion *a* lautet, falls nur eine Bahn-Drehrichtung vorkommt und die Spinrichtungen unbekannt sind,

$$a = \omega_s\, r\, e^{\pm i\,\omega\, t} \qquad \text{(Gl. 3)}$$

mit $\omega_s$ skalare Spin-Kreisfrequenz, $r$ Radius der Kreisbewegung, $\omega$ Bahn-Kreisfrequenz (ein Maß für die Geschwindigkeit), $t$ Zeit und $i = \sqrt{-1}$. Das „+" vor dem imaginären $i$ bedeutet eine rechtshändige und das „-" eine linkshändige Rotation.
Eine Rechenvorschrift ist nun vorgestellt. Aber vorm Anwenden muß noch etwas Grundsätzliches gesagt werden.

## 5. Die Trennung von Photon und Teilchen
### 5.1 Was ist ein Photon und was ist ein Teilchen?
Reine Photonen besitzen ein kugelförmiges „Herz" mit einem Bewegungsdrang. Reine Teilchen bestehen nur aus einem „Herz" und haben keinen Bewegungsdrang, sie ruhen, also sie „liegen nur dumm rum". Wir sind aber überwiegend umgeben von Mischformen zwischen Photonen und Teilchen. Und diese Mischformen sind naturgemäß langsamer als die reinen Photonen.

### 5.2 Der Heßsche Satz der Optik
In der Physikalischen Chemie hängen energetische Bilanzen nur vom Anfangs- und Endzustand ab und nicht vom Reaktionsweg. Diese Tatsache ist als Heßscher Satz bekannt. Angenommen, man interessiert sich für die molare Standard-Verbrennungsenthalpie (Brennwert) einer Reaktion, die im Labor experimentell nicht zu bewerkstelligen ist wie die Bildung von Kohlenmonoxid (CO) aus den Elementen. Das ist das einfache Beispiel, das in den Lehrbüchern der Physikalischen Chemie gewöhnlich zitiert wird. Man zerlegt die anvisierte chemische Reaktion in Teilvorgänge (auf die ich aber hier nicht eingehen möchte), deren Energiebilanz bekannt ist. Dann kann man durch geeignete Linearkombination dieser einzelnen Teilvorgänge auf den Energieinhalt einer fiktiven Reaktion schließen.

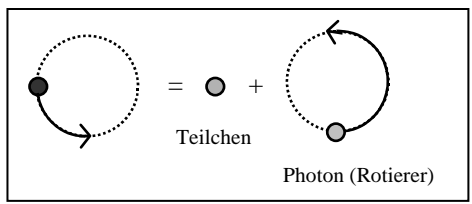

Abb. 3: Ein bewegtes Teilchen wird gedanklich in ein ruhendes Teilchen und in ein Photon aufgespalten. Die Länge der runden Pfeile symbolisiert die Bahngeschwindigkeit.

So ähnlich wollen wir mit Photonen verfahren. Photonen auf gleicher Bahn mit gleicher Bahnhelizität (Bahnhändigkeit) können sich nicht überholen und verschmelzen. Das funktioniert nur im Teilchen, weil das Teilchen wegen seiner langsameren Geschwindigkeit nicht vor dem Photon weglaufen kann. Aus Gründen der Übersichtlichkeit werden wir dennoch den fiktiven Prozeß einer reinen Photonenverschmelzung behandeln.

Aus einem propagierenden Teilchen wird das ruhende Teilchen einfach virtuell abgetrennt. Übrig bleibt dann nur noch das Photon (s. Abb. 3). Die Umkehrung dieses Vorgangs wird übrigens in Aufgabe g) behandelt. Durch die Isolierung des Photons ist die Möglichkeit eröffnet, seine Wellenfunktion aufzustellen.

In Abb. 3 und 4 stellt die Farbtiefe in den kleinen Kreisen die Masse (oder die zu ihr proportionale skalare Spin-Kreisfrequenz) symbolisch dar.

Eine weitere Anwendung des Heßschen Satzes auf die Optik ist die Möglichkeit, bei einem schwingenden Teilchen einfach die Schwingungsphotonen (Schwinger) abzutrennen, übrig bleibt dann ein ruhendes Masseteilchen (s. Abb. 4). Aber dieses Thema soll in diesem Artikel nicht weiter vertieft werden, denn es gehört in das Kapitel Phonon.

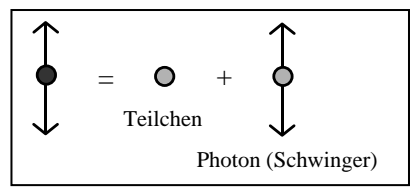

Abb. 4: Ein schwingendes Teilchen wird gedanklich in ein ruhendes Teilchen und in ein schwingendes Photon zerlegt.

## 6. Wechselwirkungen
Eine Wechselwirkung tritt ein, wenn Photonen (oder Teilchen) mit Photonen (oder Teilchen) aufeinanderstoßen (also schmerzhafte Verkehrsunfälle erleiden) und dabei verschmelzen. Dann werden die Wellenfunktionen einfach nur addiert. Sie müssen nicht normiert, geschweige denn renormiert werden.

### 6.1 Reine Spin-Spin-Wechselwirkung
Wir wollen zur Vereinfachung von der Bahnbewegung abstrahieren und nur den reinen Spin betrachten. Bei einer Spin-Spin-Wechselwirkung werden Spins gleicher oder unterschiedlicher Spinrichtung drehvektoriell addiert. Dabei bilden sich als Drehvektorsumme unterschiedliche Spin-Additionsprodukte (Herzschläge), s. Abb. 5. Alle Spins besitzen den gleichen Spin-Kreisfrequenz-Betrag $\omega_s$. In Abb. 5a bildet sich ein rotierendes „Herz" mit Spinverdopplung, in Abb. 5b ein atmendes zentrosymmetrisches „Herz" und in Abb. 5c ein atmendes achsensymmetrisches „Herz". Im zusammengeschrumpften Zustand sieht das zentrosymmetrische „Herz" punktförmig aus und das achsensymmetrische strichförmig. In Abb. 5b) und 5c) handelt es sich um zwei unterschiedliche Nullvektoren, spinlose Gebilde. (Für Zweifler möchte ich zu diesem Thema nur den Hinweis geben, im Internet bei Google nach den Stichworten „spinloses Elektron" oder „spinless electron" zu suchen.) Die Masse aller Additionsprodukte in Abb. 5 ist übrigens gleich.

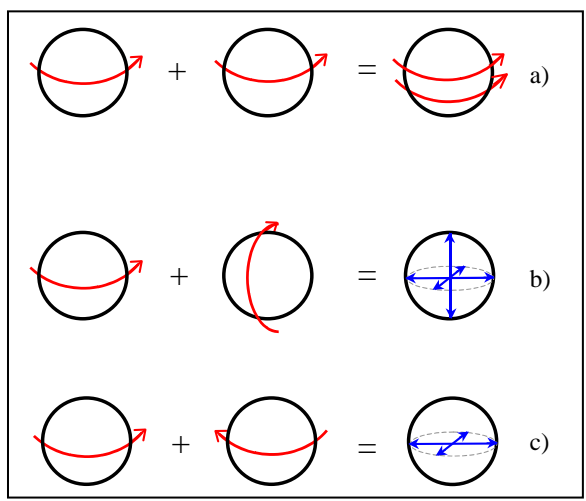

Abb. 5: Die drei Möglichkeiten einer Spin-Spin-Wechselwirkung: es verschmelzen
a) gleichsinnige Spins,
b) senkrecht aufeinander stehende Spins und
c) gegensinnige Spins.

Diese unterschiedlichen Spin-Kombinationen charakterisieren die jeweiligen magnetischen Eigenschaften der Photonen bzw. der Teilchen. Abb. 5a symbolisiert den Paramagnetismus (bis hin zum Ferromagnetismus), Abb. 5b) und 5c) zeigen zwei Formen vom Diamagnetismus. Bei der linearen Polarisation dürfte es sich um den Fall c) handeln.

### 6.2 Vereinfachte Spin-Bahn-Wechselwirkung
Die meisten vorgestellten Rechenbeispiele befassen sich mit der Wechselwirkung von Photonen. Zur Vereinfachung wird erstens die Spinrichtung vernachlässigt und zweitens wird nur eine Bahndrehrichtung um eine einzige Achse berücksichtigt. Ferner sollen nur Rotierer mit gleichem Bahnradius und demselben Kreismittelpunkt deckungsgleich verschmelzen, dies ist der Fall der totalen Durchdringung oder totalen Verschmelzung.

### 7. Wechselwirkungsbeispiele
Bei den vorgestellten Übungsbeispielen in Kapitel 7.1 agieren fast nur Photonen. Für eine Wechselwirkung zweier Photonen müssen wir folgende geometrische und sonstige Punkte kennen oder berechnen:
1. Von welcher Stelle starten die Reaktanden (in den Abbildungen 7 als schwarzer dicker Punkt symbolisiert),
2. in welche Richtung fliegen sie und
3. auf welcher Kreisbahn oder anders gearteten Bahn?
4. An welcher Stelle treffen die Reaktanden wechselwirkend aufeinander? Wo ist also der Vertex-Punkt, in den Abbildungen 7 zuweilen als Kreuz x gekennzeichnet (das x als das Symbol zweier entgegengesetzt auftreffender Pfeilspitzen).
5. Wieviel „Ladung", d. h. Spin/Masse, führen die Photonen auf ihrer Bahn mit?
6. Wie sieht nach der Wechselwirkung mit der neuen „Gewichtsverteilung" durch die skalaren Spin-Kreisfrequenzen die Form der Bahn aus?

Wir werden diese Gesichtspunkte nachstehend an einfachen Beispielen nachvollziehen.

### 7.1 Übungsbeispiele
Da wir nun einfache Beispiele betrachten wollen, können wir Gleichung 3 verwenden. Zur Orientierung werden die Achsen von Abb. 6 verwendet. Der Kreismittelpunkt der Rotierer soll im Ursprung des Koordinatensystems liegen. Wegen der Übersichtlichkeit wird bei den Aufgaben a) bis f) in den zugehörigen Abbildungen der Körper der Photonen nicht dargestellt.

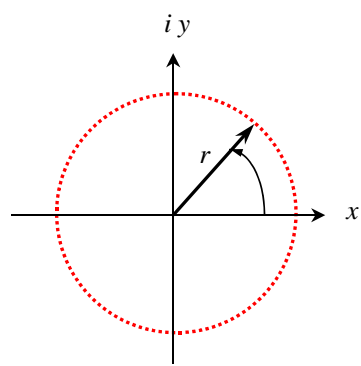

Abb. 6: Das Koordinatensystem für die Übungsaufgaben

Aufgabe a)
Zwei gleichgerichtete rechtshändige rotierende Photonen (kurz Rotierer) mit der gleichen Spin-Kreisfrequenz $\omega_s$ und dem Bahnradius $r$ sollen sich erstmals auf der x-Achse begegnen. Wie lautet das Weg-Zeit-Gesetz (die Wellenfunktion) nach der Verschmelzung?

Lösung zu Aufgabe a)
Für die Wellenfunktionen $a_0$ der beiden Rotierer gilt jeweils

$$a_0 = \omega_s \, r \, e^{i \omega t}$$

$\omega$ ist die Bahn-Kreisfrequenz der Rotierer. Die Gesamt-Wellenfunktion $a$ ergibt sich einfach durch Addition (siehe Abb. 7a).

$$a = a_0 + a_0$$

$$a = 2 \, \omega_s \, r \, e^{i \omega t}$$
===============

Die gleichsinnigen Rotierer bleiben auch nach der Verschmelzung Rotierer. Es ist so etwas wie eine „**Oberrotation**" entstanden (in Anlehnung an den Begriff Oberschwingung), indem sich das Spingewicht auf der Kreisbahn um den Faktor 2 verdoppelt hat. Damit ist also eine Spin-Kreisfrequenzverdopplung eingetreten. Dieser Effekt ist eine Möglichkeit einer Frequenzverdopplung, eine weitere wird uns in der nächsten Aufgabe begegnen.

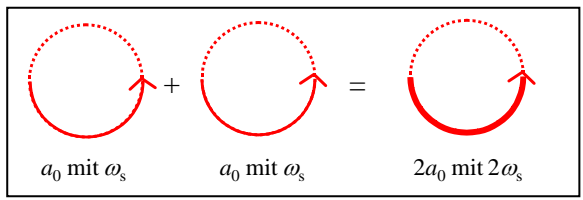

Abb. 7a: Zwei gleichgerichtete gleichfarbige Rotierer verschmelzen („Oberrotation")

Aufgabe b)
Zwei Rotierer mit der gleichen Spin-Kreisfrequenz $\omega_s$, aber mit entgegengesetzter Händigkeit sollen auf der x-Achse starten. Wie lautet das Weg-Zeit-Gesetz nach der Verschmelzung?

Lösung zu Aufgabe b)
Der Anfangspunkt wird in Abb. 7b als Punkt und der Vertex als x gekennzeichnet. Da es sich bei den Rotierern um Photonen handelt, sind sie gleich schnell. Es ist demzufolge zu erwarten, daß sie sich auf halbem Wege treffen, also ihren Vertex bei der Phase $\pi$ haben. Am Vertex lauten die Wellenfunktionen der beiden Rotierer $a_1$ und $a_2$ in der Exponentialform

$$a_1 = \omega_s\, r\, e^{i\omega t+\pi}$$
$$a_2 = \omega_s\, r\, e^{-i\omega t+\pi}$$

und in der trigonometrische Darstellung:

$a_1 = \omega_s\, r\, [\cos(\omega t + \pi) + i \sin(\omega t + \pi)]$
$a_2 = \omega_s\, r\, [\cos(\omega t + \pi) - i \sin(\omega t + \pi)]$

Die Gesamtwellenfunktion ergibt sich als Summe
$a_1+a_2 = \omega_s\, r\, [2 \cos(\omega t + \pi) + 0]$
$a_1+a_2 = -2\omega_s\, r\, \cos(\omega t)$
===============

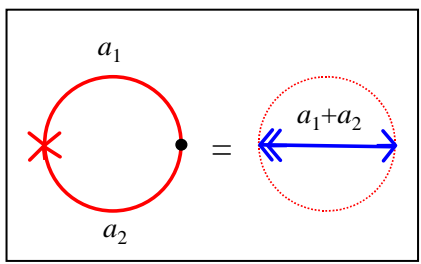

Abb. 7b: Zwei entgegengesetzte gleichfarbige Rotierer bilden einen Schwinger. (Die einfache Pfeilspitze im rechten Bild zeigt die Schwingungsrichtung in der 1. und die zweifache in der 2. Schwingungsphase an.)

Die beiden entgegengesetzten Rotierer verschmelzen also zu einem **Schwinger**, startend auf der negativen Seite der x-Achse. Auch dieses ist ein Fall einer Frequenzverdopplung, aber halt etwas anders als in Aufgabe a).

Aufgabe c)
In Analogie zu Aufgabe b) sollen die zwei entgegengesetzten Rotierer auf der y-Achse starten. Wie lautet die Wellenfunktion?

Lösung zu Aufgabe c:
Der Ausgangspunkt liegt bei $\pi/2$, nach einem weiteren Bogenweg $\pi$ erfolgt der Vertex. Dann gilt für die beiden rotierenden Photonen
$a_5 = \omega_s\, r\, [\cos(\omega t + 3\pi/2) + i \sin(\omega t + 3\pi/2)]$
$a_6 = \omega_s\, r\, [\cos(\omega t + 3\pi/2) - i \sin(\omega t + 3\pi/2)]$

$a_5+a_6 = \omega_s\, r\, [2 \cos(\omega t + 3\pi/2) + 0]$
$a_5+a_6 = 2\omega_s\, r\, \cos(\omega t + 3\pi/2)$

$a_5+a_6 = 2\omega_s\, r\, \sin(\omega t)$
===============
Die entgegengesetzten Rotierer, von der y-Achse startend, verschmelzen zu einem Sinus-Schwinger („Senkrecht-Schwinger").

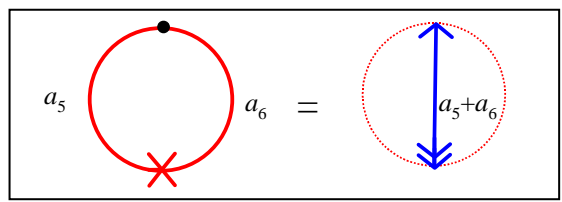

Abb. 7c: Die entgegengesetzten Rotierer, von der y-Achse startend, verschmelzen zu einem „Sinus-Schwinger". (Bild rechts: 1. Schwingungsphase eine Pfeilspitze, 2. Schwingungsphase zwei Pfeilspitzen)

Aufgabe d)
Es sollen entgegengesetzte Rotierer mit unterschiedlichen Spin-Kreisfrequenzen vereinigt werden. Das linkshändig rotierende Photon soll die doppelte Spin-Kreisfrequenz besitzen.

Lösung zu Aufgabe d)
Startpositionen und Vertex sind wie bei Aufgabe b), siehe Abbildung 7 c. Die einzelnen Wellenfunktionen lauten am Vertex

$$a_7 = \omega_s\, r\, e^{i\omega t+\pi}$$

$$a_8 = 2\omega_s\, r\, e^{-i\omega t+\pi}$$

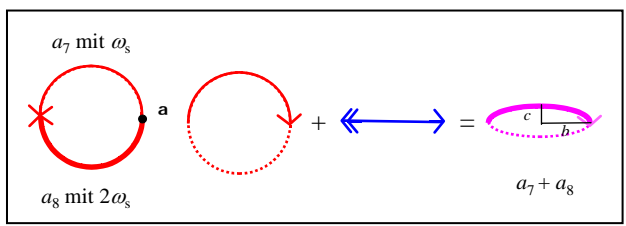

Abb. 7d: Zwei ungleichfarbige entgegengesetzte Rotierer bilden eine elliptische Rotation.

Nach den Erfahrungen der Aufgabe b) wird sich eine Schwingung auf der x-Achse ergeben mit der „Spin-Teilgewicht" $2\omega_s$, übrig bleiben wird ein Rotierer mit dem „Spin-Teilgewicht" $\omega_s$, der wie $a_8$ linkshändig ist, s. mittlerer Teil von Abbildung 7d. Die Vereinigung von Schwinger und Rotierer muß eine elliptische Bewegung ergeben.

Jetzt folgt die Berechnung aus den trigonometrischen Gleichungen:
$a_7 = \omega_s r [\cos(\omega t + \pi) + i \sin(\omega t + \pi)]$
$a_8 = 2\omega_s r [\cos(\omega t + \pi) - i \sin(\omega t + \pi)]$

Die Gesamt-Wellenfunktion ergibt sich als Summe
$a_7+a_8 = 3\omega_s r \cos(\omega t + \pi) - i \omega_s r \sin(\omega t + \pi)$
==============================
Die Gesamt-Wellenfunktion hat die Form einer Ellipsengleichung mit den Hauptachsen $b = 3\omega_s r$ und $c = \omega_s r$.

Aufgabe e)
Zwei senkrecht zueinander schwingende Photonen der gleichen Farbe sollen sich auf der x-Achse vereinigen, und zwar der „Sinus-Schwinger" im Schwerpunkt mit der „Pfeilspitze" eines „Kosinus-Schwingers". Wie lautet die Gesamt-Wellenfunktion?

Lösung zu Aufgabe e)
Die Sinus-Schwingerfunktion ist vor dem Vertex im Schwingungsmittelpunkt reell. Aber am Vertex auf der x-Achse muß man wegen der Verschmelzung der beiden Photonenkörper das imaginäre $i$ aus der „Tasche ziehen" (meinetwegen als Operator). Die Gesamt-Wellenfunktion lautet dann
$a = 2\omega_s r [\cos(\omega t) + i \sin(\omega t)]$
======================
Das ist die Wellenfunktion der Kreisbewegung.

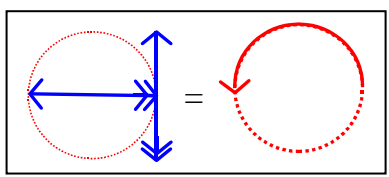

Abb. 7e: Zwei senkrecht aufeinander stehende Schwinger werden nach dem Verschmelzen auf der x-Achse ein Rotierer.

Hier folgen noch einige Erklärungen zu den Pfeilspitzen der Schwinger im linken Teil der Abbildung 7e: Waagerecht-Schwinger: 1. Schwingungsphase von 0 bis $\pi$ eine Pfeilspitze. 2. Schwingungsphase von $\pi$ bis $2\pi$ zwei Pfeilspitzen. Senkrecht-Schwinger: 1. Schwingungsphase von 0 bis $\pi/2$ eine Pfeilspitze; nächste Schwingungsphase von $\pi/2$ bis $3\pi/2$ zwei Pfeilspitzen.

Zwei senkrecht aufeinander stehende Schwinger gleicher Farbe erzeugen eine Rotierer, sofern die Pfeilspitze des einen mit dem Schwerpunkt des anderen aufeinandertreffen, also wenn die Schwinger einen Phasenunterschied von $\pi/2$ besitzen.

Nur für Kreisbewegungen muß das imaginäre $i$ in der Wellenfunktion stehen.

Aufgabe f)
Der Photonenschwinger der Aufgabe b) und ein dazu spiegelbildlicher Schwinger sollen symmetrisch aufeinandertreffen. Was passiert?

Lösung zu Aufgabe f)

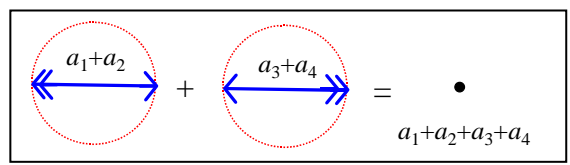

Abb. 7f: Zwei entgegengesetzte „Waagerecht-Schwinger" vereinigen sich in ihren Schwingungsmittelpunkten zu einem ruhenden Teilchen. Die Schwingungsbewegungen verschwinden. Die Summe der Spin-Kreisfrequenz-Beträge bleibt erhalten.

Die Schwinger treffen im Schwingungsmittelpunkt aufeinander, der früher der Kreismittelpunkt war. Wir schreiben die beiden Wellenfunktionen hin, die nun dort vereinigt werden sollen.

$a_1+a_2 = -2\omega_s r \cos(\omega t + \pi/2)$
$a_3+a_4 = 2\omega_s r \cos(\omega t + \pi/2)$

Die Gesamt-Wellenfunktion ist dann
$a_1+a_2+a_3+a_4 = 0$.
============
Die Schwingungsbewegungen haben sich total aufgehoben. Die Betragssumme der Gesamtspin-Wellenfunktion beträgt $4\omega_s$. Es ist also ein **ruhendes Teilchen** („Schwarzes Loch") entstanden.

Aufgabe g)
Das ruhende Teilchen der Aufgabe f) wird von einem rotierenden Photon der gleichen Spinbeladung $4\omega_s$ zentral getroffen. Was passiert? Wie lautet die Wellenfunktion?

<u>Lösung zu Aufgabe g)</u>
Die „Gesamt-Spin-Ladung" (die skalare Gesamt-Spin-Kreisfrequenz) von Teilchen und Photon beträgt $4\omega_s + 4\omega_s = 8\omega_s$ („Energiesatz"). Das Photon überträgt einen Drehimpuls auf das Teilchen derart, daß nun das Teilchen mit halber Lichtgeschwindigkeit auf der Kreisbahn fliegt. Wenn das reine Photon vorher eine Bahn-Kreisfrequenz von $\omega$ besaß, so hat das mit dem Teilchen verschmolzene Photon nur noch $\omega/2$.

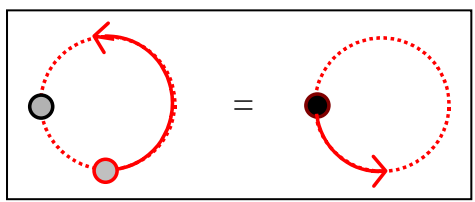

Abb. 7g: Das Photon (kleiner roter Kreisring) wird durch das ruhende Teilchen (schwarzer Kreisring) auf halbe Lichtgeschwindigkeit gebremst (kürzerer runder Pfeil im rechten Bild), bzw. das Teilchen wird durch das Photon bewegt. Das Teilchen wird schwerer (angedeutet durch die Farbtiefe im kleinen Kreisring).

Mit der halbkonventionellen Rechnung nach dem Drehimpuls-Erhaltungssatz kann die Bahngeschwindigkeit des bewegten Teilchens $v$ ermittelt werden. Dabei ist $c$ wieder die Vakuum-Lichtgeschwindigkeit. Statt der Masse wird die der Masse proportionale Größe, die Summe der Spin-Kreisfreqenz-Beträge, eingesetzt, $4\omega_s$ für das Photon und $8\omega_s$ für das bewegte Teilchen. Gemäß dem Ansatz

**Drehimpuls des Teilchens + Drehimpuls des Photons vor dem Stoß = Drehimpuls des bewegten Teilchen nach dem Stoß**

ergibt sich dann
$0 + r\, 4\omega_s \cdot c = r\, 8\omega_s \cdot v$
$v = \tfrac{1}{2} c$
Die Wellengleichung lautet dann

$$a = 8\,\omega_s\, r\, e^{i\frac{\omega}{2}t}$$

==============
Das Teilchen fährt auf einer Kreisbahn mit der halben Bahn-Kreisfrequenz des Photons.

**7.2 Abschließende Bemerkungen zu den Wechselwirkungsprodukten der Photonen**
Die vorgestellten Übungsbeispiele ähneln teilweise den einfachen Lissajous-Figuren. Ganz aus dem Rahmen fielen naturgemäß die Aufgabe a) mit der „Oberrotation" und Aufgabe f) mit der Bildung des „Schwarzen Lochs", denn hier standen die agierenden Photonen nicht senkrecht aufeinander. Bei Aufgabe a) wird übrigens gegen das Pauli-Prinzip verstoßen.

Um ein ruhendes Teilchen zu bilden, sind zwei Schwinger gleicher Farbe nötig, die entgegengesetzt getaktet sind und dank der symmetrischen Geometrie (gleiche Amplitude, Vertex in der Mitte der Schwingung) total verschmelzen. Wir haben es hier mit echter Ruhmasse zu tun. Die Bildung des ruhenden Teilchen könnte der Extremfall für das Modell des Mott-Isolators sein [8], [9].

Eine Vermutung am Rande: Möglicherweise ist dieser Vorgang nicht reversibel. Die alte Kreisbahn ist nicht automatisch wieder erreichbar und vor allem die in einem Punkt verschmolzene Spin-Masse ist nicht mehr nachträglich teilbar. Es handelt sich hier also wohl um einen irreversiblen Vorgang. Alles altert. **Photonen sind Lebewesen**, **Teilchen sind Lebewesen**.

In Aufgabe f) wird Bahn-Schwingung bzw. Geschwindigkeit vernichtet. Dieser Fall soll bei den Spin-Schwingungen der Abb. 5b) nicht passieren, da es zu ihnen keine entgegengetaktete Spin-Schwingung geben soll. Sonst würde sich Masse vernichten, was nicht vorkommen darf.

Die wichtigsten **Wechselwirkungsprodukte** der Photonen sind **Rotierer, Schwinger** und **ruhende Teilchen**. Rotierer erkennt man übrigens mathematisch daran, daß in der Wellengleichung ein „$i$" vorkommt.

Bei gleicher Spin-Kreisfrequenz benötigt der Rotierer den größten Platzbedarf im Raum, dann folgt bei gleichem Radius der Schwinger, das ruhende Teilchen beansprucht den geringsten Platz im Ortsraum. Mit anderen Worten heißt das, die Entropie sinkt in der Reihe Rotierer, Schwinger, ruhendes Teilchen. Es gibt selbstverständlich auch Mischformen zwischen den Rotierern, Schwingern und ruhenden Teilchen. In Aufgabe d) beegnete uns das Zwischending zwischen Rotierer und Schwinger, das elliptisch fliegende Photon.

Mit einer vorsichtigen Phantasie könnte man sagen, die Rotierer entsprechen den Fermionen und die Schwinger den Bosonen. Eine völlige Analogie ist nicht herzustellen, da andere physikalischen Denkmodelle zu Grunde liegen.

**8. Schluß und Ausblick**
Es konnte gezeigt werden, daß es prinzipiell möglich ist, für Photon und Teilchen eine deterministische Wellenfunktion aufzustellen. Der entscheidende Teil auf diesem Weg war wohl die Einführung der Betragssumme der Spin-Kreisfrequenzen als das Gewicht der Wellenfunktion. Quantenzahlen können dabei weggelassen werden. Bei einer Wechselwirkung müssen nur die Wellenfunktionen addiert zu werden. Es ist keine Normierung, geschweige denn Renormierung erforderlich.

Welle und Teilchen sind nun endlich identisch geworden, da sie durch das gleiche Modell erfaßt werden.

Ich hoffe hiermit, ein Modell geschaffen zu haben, das universell anwendbar ist, das die Gravitation, den Elektromagnetismus, die schwache und starke Wechselwirkung automatisch abdeckt. Mit der modernen Lasertechnologie ist es immerhin möglich, in die Gefilde der Hochenergiephysik zu stoßen.

Achtung, man darf es mit der Determiniertheit nicht übertreiben. In vielen Fällen sind nicht alle Bedingungen bekannt, um die vollständige Determiniertheit in der Praxis um jeden Preis realisieren zu können. Man ist nach wie vor immer noch auf geeignete statistische Methoden angewiesen. Sorgen macht mir z. B. die Größe des Radius´ der Kreisbewegungen.

**9. Literatur**

Sämtliche Weblinks wurden zuletzt getestet am 26.09.2006.

[1] SCHULZ, P.: Kurioses zum Photon: In: *DPG-Vortragsband Didaktik der Physik*, Berlin 1997, S. 624-629

[2] SCHULZ, P.: Plausible Definition von Masse, Ladung und Spin. In: *CD zur Frühjahrstagung Didaktik der Physik in der Deutschen Physikalischen Gesellschaft,* Dresden 2000
http://home.arcor.de/gruppederneuen/Seiten/Publikationen/DefinitionMasseLadung.pdf

[3] SCHULZ, P.: Plausible Erklärungshinweise gegen die Überlichtgeschwindigkeit. In: *CD zur Frühjahrstagung Didaktik der Physik in der Deutschen Physikalischen Gesellschaft*, Bremen 2001
http://de.arxiv.org/abs/physics/0609221
http://de.arxiv.org/ftp/physics/papers/0609/0609221.pdf

[4] SCHULZ, P.: Infrarotspektroskopie plausibel. In: *CD zur Frühjahrstagung des Fachverbandes Didaktik der Physik in der Deutschen Physikalischen Gesellschaft*, Leipzig 2002
http://de.arxiv.org/abs/physics/0307062
http://de.arxiv.org/ftp/physics/papers/0307/0307062.pdf
SCHULZ, P.: Infrared Spectroscopy Plausible. In: Progress in Chemical Physics Research. Editor: Linke, A. N., S. 121-135. New York: Nova Science Publishers, Inc. 2005. - ISBN 1-59454-451-4

[5] SCHULZ, P.: Photon plausibel bei linearer Polarisation. In: *CD zur Frühjahrstagung des Fachverbandes Didaktik der Physik in der Deutschen Physikalischen Gesellschaft*, Augsburg 2003. – ISBN 3-936427-71-2
http://de.arxiv.org/abs/physics/0609192
http://de.arxiv.org/ftp/physics/papers/0609/0609192.pdf

[6] Kuhn, A.: Einzelne Atome im Resonator. In: *Physik in unserer Zeit*, 37 (2006), S. 6

[7] BOARDMAN, A. D.; MARINOV; K.; ZHELUDEV, N; FEDOTOV, V. A.: Nonradiating toroidal structures. In:
http://arxiv.org/abs/physics/0510155
http://arxiv.org/ftp/physics/papers/0510/0510155.pdf

[8] GREINER, M.; MANDEL, O.; ESSLINGER, E.; HÄNSCH, T. W.; BLOCH, I.: Quantum phase transition From a superfluid to a Mott insulator in a gas of ultracold atoms. In: *Nature* 415, S. 39-44 (2002)

[9] GREINER, M.; MANDEL, O.; HÄNSCH, T. W.; BLOCH, I: Collaps and revieval of the matter wave of a Bose-Einstein condensate. *Nature* 419, S. 51-54 (2002)

## 10. Anhang
### 10.1 Mathematische Erläuterung zum Bahndrehimpuls-Modell

Alle Bewegungen erfolgen auf Kreisbahnen. Ich zergliedere die Rotationsbewegung in solche um die Achsen eines kartesischen Koordinatensystems mit den linearen Achsrichtungen $x$, $y$ und $z$. Der Weg-Zeit-Vektor $\boldsymbol{a}$ hat die drei axialen Komponenten $a_x$, $a_y$ und $a_z$ (siehe Abb. 1). Die Indizes deuten auf die Achse hin, um die sich das Photon dreht.

$$\boldsymbol{a} = (a_x, a_y, a_z) \quad (Gl.\ 1)$$

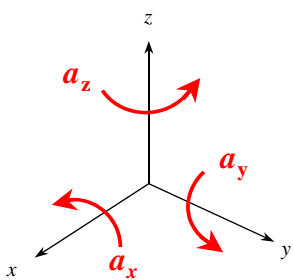

Abb. 1: Die drei axialen Komponenten in einem rechtshändigen kartesischen Koordinatensystem

Ich benutze den Formalismus der komplexen Zahlen und erweitere vom ebenen Raum der Abb. 2 bzw. vom eindimensionalen auf den dreidimensionalen Drehvektorraum. Das imaginäre $i$

$$i = \pm\sqrt{-1}$$

wird aus formalen Gründen mit den zugehörigen Achsen indiziert. Die indizierten $i$'s entstehen durch Multiplikation der Einheitsvektoren von der entsprechenden x-, y-, z-Achse ($e_x$, $e_y$, $e_z$) mit $i$:

$$i_x = i\, e_x$$
$$i_y = i\, e_y$$
$$i_z = i\, e_z$$

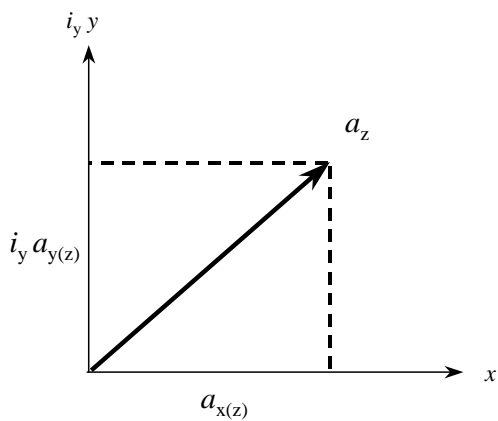

Abb. 2: Die beiden polaren Komponenten einer komplexen Zahl $a_z$ in der xy-Ebene

Betrachten wir exemplarisch die Drehung um die z-Achse, siehe Abb. 2. Dann befinden sich die polaren Teilvektoren $a_{x(z)}$ und $a_{y(z)}$ in der xy-Ebene. Sie spannen die Komponenten auf für die Drehimpuls-Vektorkomponente $a_z$. Der erste Index bei den linearen (also polaren) Teilvektoren gibt die jeweilige Achsenrichtung der Teilschwingung an, in der Klammer steht die Drehrichtung der axialen Komponente.

Die <u>Vektorform</u> lautet dann

$$a_z = a_{x(z)} \pm i_y\, a_{y(z)}. \quad (Gl.\ 2)$$

Das $\pm$ vor dem imaginären $i_y$ hat folgende Bedeutung: Für rechtshändige Drehungen ist das positive Vorzeichen anzusetzen und für linkshändige Drehungen das negative.

Jetzt folgt die <u>trigonometrische Form</u> (s. Abb. 3):

$$a_z = r_z\,[\cos(\varphi_z + 2\pi k_z) \pm i_y \sin(\varphi_z + 2\pi k_z)] \quad (Gl.\ 3)$$

mit $k_z = 0, \pm 1, \pm 2, ...$

Die Größe $r_z$ ist der Betrag der komplexen Zahl. Das Argument $\varphi_z$ ist der Kreisbogen, gemessen von der x-Achse im Rechtsschraubensinn (also im entgegengesetzten Uhrzeigersinn). Und $k_z$ ist ein Maß für die Phase. Der Index z bezeichnet für alle vorkommenden Größen die Drehung um die z-Achse.

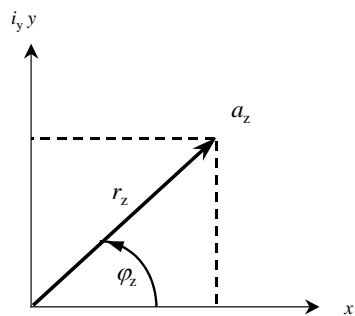

Abb. 3: Die notwendigen Parameter einer komplexen Zahl $a_z$ für die trigonometrische Form

Die <u>Exponentialform</u> lautet:

$$a_z = r_z\, e^{\pm i_y(\varphi_z + 2\pi k_z)}. \quad (Gl.\ 4)$$

### Das Weg-Zeit-Gesetz

Um zu einem Weg-Zeit-Gesetz zu gelangen, muß der durchfahrene Kreisbogen $\varphi_z$ durch die Kreisfrequenz $\omega_z$ mal Zeit $t$ ersetzt werden, s. Gleichung 5:

$$\varphi_z = \omega_z\, t. \quad (Gl.\ 5)$$

Den Phasenbeitrag $2\pi k_z$ setze ich erst einmal zu Null, um dem Schreibwust zu entkommen und um die Wechselwirkungen an exemplarischen Beispie-len mit physikalischem Hintergrund einfacher beschreiben und gegebenenfalls frisch überdenken zu können.

$\omega_z$ ist die Kreisfrequenz des Photons (oder Teilchens) auf dem Radius $r_z$ um die z-Achse. Dann erhält man durch Einsetzen von Gleichung 5 in Gleichung 4 die Gleichung 6 als Weg-Zeit-Gesetz aus lauter physikalischen Meßgrößen.

$$a_z = r_z\, e^{\pm i_y\, \omega_z t} \quad \text{(Gl. 6)}$$

Nochmals sei an dieser Stelle wiederholt: Das "+" vor dem imaginären $i_y$ bedeutet eine rechtshändige und das "-" eine linkshändige Umdrehung. Solange keine Wechselwirkung stattfindet, sind $r_z$ und $\omega_z$ konstante Größen, nur die Zeit $t$ fließt gleichförmig dahin, wächst also linear an.

Sämtliche Formeln dieses Kapitels werden am Schluß zusammengestellt und darüber hinaus auch die Gleichungen für die Drehungen um die x- und y-Achse ergänzt.

$r_z$ hat folgende physikalische Bedeutung, die hier ebenfalls noch einmal wiederholt wird: Es ist der **Radius der Drehung** um die z-Achse. Später – im Kapitel Wechselwirkungen - muß die Deutung auf Schwingungen erweitert werden. $r_z$ ist dann die **Amplitude der Schwingung**. Das Argument $\varphi_z$ ist der Kreisbogen, der von der x-Achse startend, im Rechtsschraubensinn umfahren wird. Der Analogieschluß von Gleichung 6 auf Gleichung 4 ist der folgende: $\varphi_z$ ist der Kreisbogen, der von der x-Achse im Rechtsschraubensinn ausgehend, in einem geeignet gewählten Zeitraum umrundet wird, das Ganze in regelmäßigen Abständen durch ein schmales Zeitfenster betrachtet. Gleichung 4 kann diese Momentaufnahme optisch wiedergeben, siehe Abb. 3.

Durch die Größe Zeit $t$ ist prinzipiell die Determiniertheit in der Physik automatisch eingebaut worden, wie wir später im Kapitel Wechselwirkung sehen werden. $\omega_z$ ist übrigens die einzig mögliche Bahn-Kreisfrequenz des Photons auf dem Radius $r_z$.

Wenn ich an dieser Stelle abbrechen würde, hätte ich einen ganz schlimmen Fehler begangen, der sich nicht wieder in der Physik abspielen darf. Es darf keine kinematische Physik zementiert werden, die Physik ohne Masse (und ohne Spin).

### 10.2 Anmerkung

Zu dem kinematischen Teil des Bahndrehimpuls-Modells möchte ich folgendes anmerken: Kurz nachdem ich mein Manuskript zum Vortrag fertiggestellt hatte, konnte ich mir unter dem Begriff „Quaternionen" nichts vorstellen. Nur das Wort hatte ich schon einmal gehört. Rein zufällig fand ich bei der Google-Suche das Schlagwort „Quaternionen" und konnte Einzelheiten im Online-Lexikon Wikipedia nachlesen. So mußte ich also feststellen, daß ich unabsichtlich das Modell der Quaternionen neu erfunden habe. (Maxwell hat übrigens seine Gleichungen 1873 in der Quaternionen-Schreibweise veröffentlicht.)

Hätte ich allerdings vorher schon das Thema Quaternionen gekannt, wäre ich niemals bis zur Formulierung einer deterministischen Wellenfunktion vorgedrungen.

### 10.3 Formelsammlung
Jetzt folgt eine Zusammenstellung der Formeln für komplexe Zahlen zum Dreh-Vektormodell für alle Achsen am Beispiel rechtshändiger Drehungen:

- **Kinematischer Teil**

| Drehung um Achse | Ebene | **Vektorform** s. Abb. 2 |
|---|---|---|
| x | yz | $a_x = a_{y(x)} + i_z\, a_{z(x)}$ |
| y | zx | $a_y = a_{z(y)} + i_x\, a_{x(y)}$ |
| z | xy | $a_z = a_{x(z)} + i_y\, a_{y(z)}$ |

| Drehung um Achse | Ebene | **Trigonometrische Form** s. Abb. 3 |
|---|---|---|
| x | yz | $a_x = r_x\,[\cos(\varphi_x + 2\pi k_x) + i_z \sin(\varphi_x + 2\pi k_x)]$ |
| y | zx | $a_y = r_y\,[\cos(\varphi_y + 2\pi k_y) + i_x \sin(\varphi_y + 2\pi k_y)]$ |
| z | xy | $a_z = r_z\,[\cos(\varphi_z + 2\pi k_z) + i_y \sin(\varphi_z + 2\pi k_z)]$ |

**Exponentialform**

---

Drehung um x-Achse in yz-Ebene:

$$a_x = r_x\, e^{+i_z(\varphi_x + 2\pi k_x)}$$

---

Drehung um y-Achse in zx-Ebene:

$$a_y = r_y\, e^{+i_x(\varphi_y + 2\pi k_y)}$$

---

Drehung um z-Achse in xy-Ebene:

$$a_z = r_z\, e^{+i_y(\varphi_z + 2\pi k_z)}$$

---

- **Deterministische Wellenfunktion**

In dem folgenden Abschnitt sind die Überlegungen bezüglich der Spin-Kreisfrequenz aus den Kapiteln 3 und 4 eingeflossen.

---

Drehung um x-Achse in yz-Ebene:

$$a_x = \omega_s\, r_x\, e^{+i_z(\omega_x t + 2\pi k_x)}$$

---

Drehung um y-Achse in zx-Ebene:

$$a_y = \omega_s\, r_y\, e^{+i_x(\omega_y t + 2\pi k_y)}$$

---

Drehung um z-Achse in xy-Ebene:

$$a_z = \omega_s\, r_z\, e^{+i_y(\omega_z t + 2\pi k_z)}$$

---